\documentclass[amsmath,amssymb,amsbsy,reprint,pra,preprintnumbers,showpacs,superscriptaddress]{revtex4-1}
\usepackage{graphicx,color}
\usepackage{dcolumn}
\usepackage{bm}
\usepackage{braket}
\usepackage{mathtools}
\usepackage{ulem}
\usepackage[breaklinks,colorlinks=true,linkcolor=blue,urlcolor=blue,citecolor=blue]{hyperref}
\usepackage{times}
\usepackage{physics}
\usepackage{latexsym}
\usepackage{amsmath, amssymb}
\usepackage{mathtools}
\usepackage{multirow}

\newcommand{\be}{\begin{eqnarray}}
\newcommand{\ee}{\end{eqnarray}}

\begin{document}

\title{Measurement-only dynamical phase transition of topological and boundary order\\
in toric code and gauge-Higgs models
}
\date{\today}
\author{Takahiro Orito}
\affiliation{Institute for Solid State Physics, The University of Tokyo, Kashiwa, Chiba, 277-8581, Japan}

\author{Yoshihito Kuno} 
\affiliation{Graduate School of Engineering Science, Akita University, Akita 010-8502, Japan}
\author{Ikuo Ichinose} 
\affiliation{Department of Applied Physics, Nagoya Institute of Technology, Nagoya, 466-8555, Japan}


\begin{abstract}
We extensively study long-time dynamics and fate of topologically-ordered state in toric code model  evolving through projective measurement-only circuit. 
The circuit is composed of several measurement operators corresponding to each term of toric code Hamiltonian with magnetic-field perturbations, which is a gauge-fixed version of a (2+1)-dimensional gauge-Higgs model.
We employ a cylinder geometry with distinct upper and lower boundaries to classify stationary states after long-time measurement dynamics. 
The appearing stationary states depend on measurement probabilities for each measurement operator. 
The Higgs, confined and deconfined phases emerge in the time evolution by the circuit. 
We find that both Higgs and confined phases are separated from the deconfined phase by topological entanglement entropy.
We numerically clarify that both Higgs and confined phases are characterized by a long-range order on the boundaries accompanying edge modes, and they shift with each other by a crossover reflecting properties in the bulk phase diagram.
\end{abstract}


\maketitle
\section{Introduction} 
Projective measurements performed on quantum circuit generate specific dynamics and produce exotic quantum many-body states. 
As a recent exciting phenomenon, measurement-induced entanglement phase transition in quantum circuit is attracting attention of broad physicist communities, in which phase transition takes place by the interplay and competition between projective measurements and unitary gates \cite{Li2018,Skinner2019,Li2019,Vasseur2019,Chan2019,Szyniszewski2019,Choi2020,Bao2020,Gullans2020,Jian2020,Zabalo2020,Sang2021,Sang2021_v2,Nahum2021,Sharma2022,Fisher2022_rev,Block2022,Liu,Richter2023,Sierant2023,Kumer2023}.

Similar measurement-induced phase transitions are also observed in {\it measurement-only circuit} (MoC). With suitable choice of measurement operations and their frequencies, MoC generates non-trivial phases of matter; symmetry protected topological (SPT) phases \cite{Lavasani2021,Klocke2022,KI2023}, topological orders \cite{Lavasani2021_2,Negari2023}, 
and non-trivial thermal and critical phases \cite{Ippoliti2021,Sriram2023,KOI2023,Lavasani2023,Zhu2023}. 
These dynamical phenomena originate from non-commutativity and back action of measurements, and the interplay of them produces intriguing stationary states after `time evolution' through circuit. 
Furthermore, suitable measurement on initially-prepared states can produce resource states for quantum computation \cite{Raussendorf2001,Briegel2009}, etc.
Example of this ability includes production of  long-range entanglement states with intrinsic topological order from simple SPT states \cite{Tantivasadakarn2022}.

A recent experiment \cite{Google_quantum_AI} has realized toric code \cite{Kitaev2003}. The robustness of the system to decoherence error, measurements or magnetic perturbation is now open and one of the attracted issues \cite{J-Y-Lee2023,Fan2023,Wang2023}.

With the change of geometry of system, how the property of the system changes is an essential issue. 
For example, in theoretical level, introduction of the cylinder geometry with rough boundary and combination of various types of measurements can exhibit rich physical phenomena to the toric code system \cite{Verresen2022,Wildeboer2022,Negari2023}.
Furthermore, there are fascinating findings obtained by interplay between the viewpoint of lattice gauge theory \cite{Kogut1979} and the notion of SPT phases \cite{Verresen2022}.
That is, for the gauge-Higgs model, the boundary state of the cylinder geometry exhibits a kind of long-range order (LRO)  with spontaneous symmetry breaking (SSB), and through the SSB pattern, the Higgs and confined phase in the model are to be distinguished, contrary to the common belief that the two phases are adiabatically connected without any transition singularities \cite{Fradkin1979}. 

The above discussion was obtained by the Hamiltonian formalism. 
In this paper, we shall study circuit dynamics of toric code system with the cylinder geometry by using measurement-only dynamics, where we employ distinct upper and lower boundaries, elucidating a duality between the Higgs and confined phases. This gives an efficient setup for further investigation of the behavior of the transition between the Higgs and confined phases.
In particular, we study how initially-prepared toric code state (topological ordered state) dynamically changes.
In this process, the choice of measurement operators and their probability are essential ingredients.

\begin{figure*}[t]
\begin{center} 
\includegraphics[width=17cm]{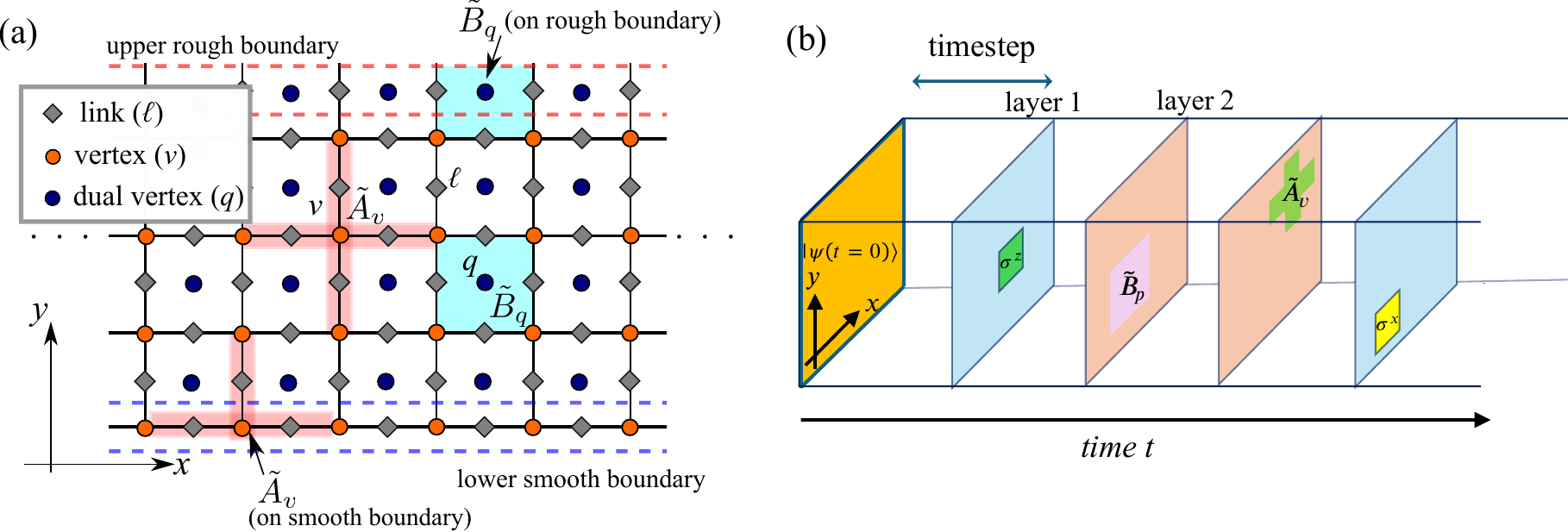}  
\end{center} 
\caption{(a) Schematic figure of the lattice system on cylinder geometry with upper rough and lower smooth boundaries. 
The star operator (the red-shaded object) and the plaquette operator (the right-blue-shaded object) are considered. 
The red and blue dashed regimes are upper rough and lower smooth boundaries, respectively. 
(b) Schematic image of the measurement-only circuit. 
The initial stabilizer state denoted by $|\psi(t=0)\rangle$ is the exact unique ground state of  toric code on the cylinder geometry with the boundaries shown in (a).}
\label{Fig1}
\end{figure*}
We shall show that rich dynamical behavior and stationary states emerge in the system by varying protocol of the MoC.
In particular, we find that Higgs and confined stationary states are separated via crossover behavior under this geometry, and then they are observed in the MoC by  
the boundary orders, in contrast to the well-known observation in \cite{Fradkin1979}. 
According to our previous work \cite{KI2023_2}, we expect that the obtained result gives an important insight into the \textit{boundary physics}
of the corresponding Hamiltonian system, for which only qualitative conjecture has been given so far.
Furthermore, when we apply strictly competitive protocol for projective measurements, the saturation time gets very long, indicating that competitive dynamics takes place in critical regime among deconfined, Higgs and confined phases. 
On the other hand, we also observe that topological order (TO) of the initial toric code under this specific boundary conditions is sustained by frequent measurement of the toric code stabilizers as a 
result of reduction of the projective-measurement effects.
This dynamical process can be useful for quantum error correction \cite{Botzung}.

The rest of this paper is organized as follows. 
In Sec.~II, we explain the setup of MoCs, where two different measurement layers are introduced. 
Stabilizer generators of the stabilizer formalism, which we employ for the MoC, depend on the cylinder geometry as well as open boundary conditions; upper rough and lower smooth boundaries are used in order
to investigate the Higgs and confined phases simultaneously.
To clarify the aim of the present study,
we also introduce and explain a reference Hamiltonian to the dynamics of this MoC. 
In the first layer, decoherent single-site (i.e., single-qubit) measurement is performed hindering the TO, whereas  the second layer works as recovery of the initial TC state.
In Sec.~III, We show the practical numerical methods and physical observables to investigate properties
of emergent states and their phases.
In Sec.~IV, we give numerical results of a spin-glass (SG) order of boundary states and its relation to
an SPT.
Section V displays a numerical study of the bulk TO by employing topological entanglement entropy (TEE).
By increasing single-site measurement probability, the initial TC state tends to lose its bulk TO.
On the other hand, the boundary SG order emerges.
Section VI is devoted to conclusion and discussion.


\section{Measurement-only circuit and comparable Hamiltonian formalism}
Let us consider a lattice system composed of $L_x \times L_y$ plaquettes ($q$-lattice) and $L_x \times (L_y-1)$ vortices ($v$-lattice).
Physical qubits reside on links of the $v$-lattice. 
The total qubit number is $L\equiv 2L_xL_y$. 
We employ cylinder geometry, in which we set the upper rough and lower smooth boundaries in the $y$ direction as shown in Fig.~\ref{Fig1}(a). 

We shall investigate `time evolution' of the pure stabilizer state specified by the following set of stabilizer generators \cite{Nielsen_Chuang},  
$S_{\rm int}=\{\tilde{A}_v | v\in \mbox{all } v \} + \{ \tilde{B}_q|q\in 
\mbox{all } q \}$, 
where $\tilde{A}_v$ and $\tilde{B}_q$ are the star and plaquette operators of the toric code, defined by $\tilde{A}_v=\prod_{\ell_{v} \in v} \sigma^x_{\ell_{v}}$ and $\tilde{B}_q=\prod_{\ell_{q} \in q} \sigma^z_{\ell_{q}}$, 
with $\ell_{v} \in v$ standing for links emanating from vertex $v$, and $\ell_{q} \in q$ for links composing plaquette $q$. 
The above stabilizer generators are all linearly independent and specify the stabilizer state
$S_{\rm int}$, which is nothing but {\it the exact unique gapped ground state} of the toric code Hamiltonian without magnetic perturbations on the cylinder geometry \cite{degene1}. 
The motivation to select this geometry and unique pure states as an initial state is that we can observe some duality between the upper rough and lower smooth boundary that can elucidate the Higgs and confined phase, while we expect the bulk physics of stationary states is not significantly affected by the geometry and choice of initial states.
Throughout this work, we use the same notation for the stabilizer set and the corresponding state interchangeably.

For the stabilizer state of this lattice system, we apply sequential projective measurements as MoC. 
In the protocol, we introduce two distinct layers: in the first one (called ``layer 1"), two projective measurements with observables, 
$\hat{M}^1_{\ell}=\sigma^x_{\ell}$ and $\hat{M}^2_{\ell}=\sigma^z_{\ell}$ are applied with a uniform probability $p_x$ and $p_z$, respectively, for each link except the bottom smooth boundary (in the case of $\hat{M}^1_{\ell}$) and each link except the top rough boundary (in the case of $\hat{M}^2_{\ell}$). 
We choose these measurement points
inspired by the Hamiltonian we later commented on Eq.~(\ref{HTC}).

In the second one (called ``layer 2"), two projective measurements with observables, $\hat{R}^1_{v}=A_v$ and $\hat{R}^2_{p}=B_{p}$ are applied with the same uniform probabilities $0.5$ for each vertex and plaquette.
We consider the following measurement protocol: (I) we choose layer 1 and 2 with probability $1-p_s$ and $p_s$, and (II-a) in the case of layer 1, we perform measurement $\hat{M}^1_{\ell}$ and $\hat{M}^2_{\ell}$ with probability $p_x$ and $p_z$, respectively, or (II-b) in the case of layer 2, we perform measurement $\hat{R}^1_{v}$ and $\hat{R}^2_{p}$ with probability $0.5$. 
The schematic image of this MoC is shown in Fig.~\ref{Fig1} (b). 
One consecutive application of layers 1 or 2 corresponds to the unit of time.

Dynamics of the MoC and its stationary stabilizer states, if exist, can be inferred from the following Hamiltonian of the toric code model with open boundary conditions and in magnetic fields, 
\begin{eqnarray}
H_{\rm TC}&=&-\sum_{v}h_x\tilde{A}_v-\sum_{q}h_z\tilde{B}_q\nonumber\\
&&-\sum_{\ell\notin \mbox{smooth}}J_x \sigma^x_{\ell}
-\sum_{\ell \notin \mbox{rough}}J_z\sigma^z_{\ell}.
\label{HTC}
\end{eqnarray}
The above model is a gauge-fixing version of a lattice gauge-Higgs model with open boundary conditions employed for investigation on its topological properties~\cite{KOI2024}. 
In this work, we assume $h_{x(z)}$, $J_{x(z)} \ge 0$.
The gauge-Higgs model on infinite system was analyzed in Fradkin and Shenker \cite{Fradkin1979}, where the phase diagram with Higgs, confined and deconfined (toric code) regimes was discovered. 
$H_{\rm TC}$ in Eq.~(\ref{HTC}) for $J_x=J_z=0$ with general boundary conditions is the fixed point Hamiltonian of the deconfined phase, and its ground states are a gapped topological state called toric code.
Gauge-Higgs model related to the above TC model was recently re-investigated from viewpoint of SPT  \cite{Verresen2022}, by employing specific boundary conditions. 
In this work, we extend the above study by introducing the cylinder geometry with both rough and smooth boundaries as in Eq.~(\ref{HTC}).
Then,
the Higgs and confined phases in (2+1) D can be distinguished with each other by observing the long-range order (LRO) on the boundaries, which is a characteristic signature implying that both the Higgs and confined phases are SPTs being protected by magnetic(electric)-one-form symmetry, respectively.
Findings in this present work are, therefore, complementary to the study in Ref.~\cite{Verresen2022}.

\begin{figure}[t]
\begin{center} 
\includegraphics[width=6cm]{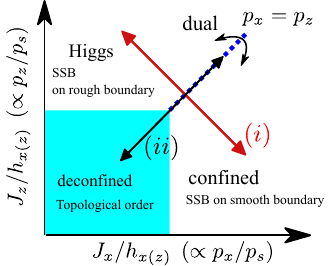}  
\end{center} 
\caption{Schematic phase diagram of the Hamiltonian $H_{\rm TC}$ in Eq.~(\ref{HTC}). 
Here, we fix $h_x=h_z$ to a finite value. 
By imposing a specific boundary conditions for a system with the cylinder geometry, the Higgs and confined phases are distinguished. The label ``SSB'' stands for spontaneous symmetry breaking. We expect the relationship between the parameter ratio and the probability ratio, $J_x/J_z\longleftrightarrow p_x/p_z$ and $h_{x(z)}/J_{x(z)}\longleftrightarrow p_s/p_{x(z)}$, which is a good guide to expect the steady state in the MoC. The blue dotted line is expected to be $p_x=p_z$ case in our MoC. Double-headed arrows (i) and (ii) represent the parameter spaces we focus on in this study. In particular, we focus on (i) in Sec.~IV and (ii) in Sec.~V.
}
\label{Fig2}
\end{figure}

It is verified by the practical calculation \cite{KI2023_2} that the gauge-Higgs and TC models are exactly connected by the gauge fixing (so called unitary gauge) even with the specific open boundary conditions employed in Ref.~\cite{Verresen2022} and also in the present study. This is quite natural as local gauge transformation can be performed at each site where the matter field resides on.
(For more detailed discussion, in particular for the bulk-edge entanglement, see Sec.~III F of Ref.~\cite{KI2023_2}.)
Therefore for the Hamiltonian of Eq.(\ref{HTC}), we expect a phase diagram such as: Higgs phase for $J_z> J_x,h_x,h_z$, confined phase for $J_x> J_z,h_x,h_z$  and deconfined (toric code) phase for $h_x=h_z>J_z, J_x$ \cite{Trebst2007,Vidal2009,Tupitsyn2010,Dusuel2011,Wu2012}. 
The schematic image of the phase diagram is shown in Fig.~\ref{Fig2}, where the schematic is based on the global phase structure of the TC model with magnetic fields under periodic boundary, which has been clarified by using Monte Carlo numerical calculation \cite{Tupitsyn2010}.
This phase diagram of the Hamiltonian of Eq.~(\ref{HTC}) gives us insights into the appearance of stationary states in the MoC started with the stabilizer state $S_{\rm int}$.

In the previous study \cite{KI2023}, we investigated relationship between Hamiltonian systems and MoCs. A brief explanation is shown in Appendix A. The results obtained there indicate the following relationship between the parameter ratios in the present system, $J_x/J_z\longleftrightarrow p_x/p_z$ and $h_{x(z)}/J_{x(z)}\longleftrightarrow p_s/p_{x(z)}$.

In the recent understanding of the above phase diagram of the gauge-Higgs model, which is an ancestor model of $H_{\rm TC}$ with exactly the same energy eigenstates,
higher-form-symmetry plays an important role \cite{Gaiotto2015,McGreevy2022}.
For the case of $J_x=0$ in $H_{\rm TC}$ in Eq.~(\ref{HTC}), there emerges magnetic-form-symmetry generated by 
the operator $W^c_\gamma\equiv \prod_{\ell\in \gamma}\sigma^z_\ell$, where $\gamma$ is an arbitrary closed loop on links of the $v$-lattice.
Similarly for a string \textit{connecting two different links residing on the rough boundary}, $\Gamma$,
$W^o_\Gamma\equiv \prod_{\ell\in \Gamma}\sigma^z_\ell$ is another one-form-symmetry (See Appendix B).
These one-form-symmetries can be robust for a finite value of $J_x$ as long as the state belongs to the same phase with that of $J_x=0$.
More precisely, the deconfined (toric code) phase is characterized as an SSB phase of the $W^c_\gamma$ and $W^o_\Gamma$
symmetries.
On the other hand, the Higgs phase is an SPT of the above two symmetries \cite{Verresen2022}.
For $J_z=0$, parallel discussion works for the confined phase as a duality picture and the electric-form-symmetries, $H^c_{\tilde{\gamma}}\equiv \prod_{\ell\in \tilde{\gamma}}\sigma^x_\ell$ and $H^o_{\tilde{\Gamma}}\equiv \prod_{\ell\in \tilde{\Gamma}}\sigma^x_\ell$, where links $\ell$'s are crossed by $\tilde{\gamma}$ and $\tilde{\Gamma}$, a close loop and string (connecting two links on the smooth boundary) on the dual lattice, respectively.

An interesting issue is whether the SPT property of the Higgs phase in the gauge-Higgs model 
is preserved in the TC model obtained from the gauge-Higgs model through the gauge fixing. 
An important observation concerning this issue is that `logical operators' such as
$W^o_\Gamma\equiv \prod_{\ell\in \Gamma}\sigma^z_\ell$ with $\Gamma$ embedded in the smooth boundary,
which play an essentially important role for edge-mode physics on the boundaries, are invariant under the gauge fixing as well as related unitary transformation~\cite{KI2023_2,Wildeboer2022}.
This is quite natural as these are gauge-invariant operators in the gauge-Higgs model.
The SSB of symmetries produced by the `logical operators' is evidence of the existence of SPT, which we verify in the present study.
In other words, one-form symmetries in the gauge-Higgs and TC models play essentially the same role for the SPT properties. 
Simply put, it is plausible to expect that Higgs=SPT is a gauge-invariant property if it exists. 

Finally, an important remark concerning the phase diagram in Fig.~\ref{Fig2} is in order.
How the boundary physics looks like in the Higgs-confined `critical regime' is an interesting unsolved problem, even though a schematic line separating the Higgs and confined phases is drawn in a 
phase diagram in Ref.~\cite{Verresen2022}.
In the previous work studying the boundary physics in the critical regime of the \textit{deconfined-Higgs} phase transition~\cite{KOI2024}, we showed that a genuine transition-like behavior emerges on the boundary 
as a result of the strict bulk phase transition.
On the other hand as the Higgs and confined phases are connected in an adiabatic matter without any singular behaviors,
it is important to investigate the boundary physics in that regime.
We shall study this problem in this work to obtain an interesting result.
We expect that the result obtained by the present MoC gives an important insight into the ground-state phase diagram of the corresponding Hamiltonian system~\cite{KI2023}.

\section{Numerical stabilizer simulation}
The projective measurement in the MoC is implemented by the stabilizer formalism \cite{Gottesman1997,Aaronson2004}. 
For layer 1; $\sigma^{z}$-projective measurement at link $\ell$ removes one of the star operators $\tilde{A}_v$ residing on the boundary vertices of $\ell$ ($v_1$ and $v_2$), and then $\sigma^{z}_\ell$ becomes a stabilizer generator as well as the product of the star operators $\tilde{A}_{v_1}\tilde{A}_{v_2}$. 
That is, the initial stabilizer state $S_{\rm int}$ tends to lose $\tilde{A}_v$'s with the probability $p_{z}$, and the number of $\sigma^{z}$-stabilizer generator increases instead. 
Thus, this process leads to decay of the initial TO. 
Similar process for $\sigma^x$ and $\tilde{B}_p$ with probability $p_x$.

\begin{figure*}[t]
\begin{center} 
\includegraphics[width=15.5cm]{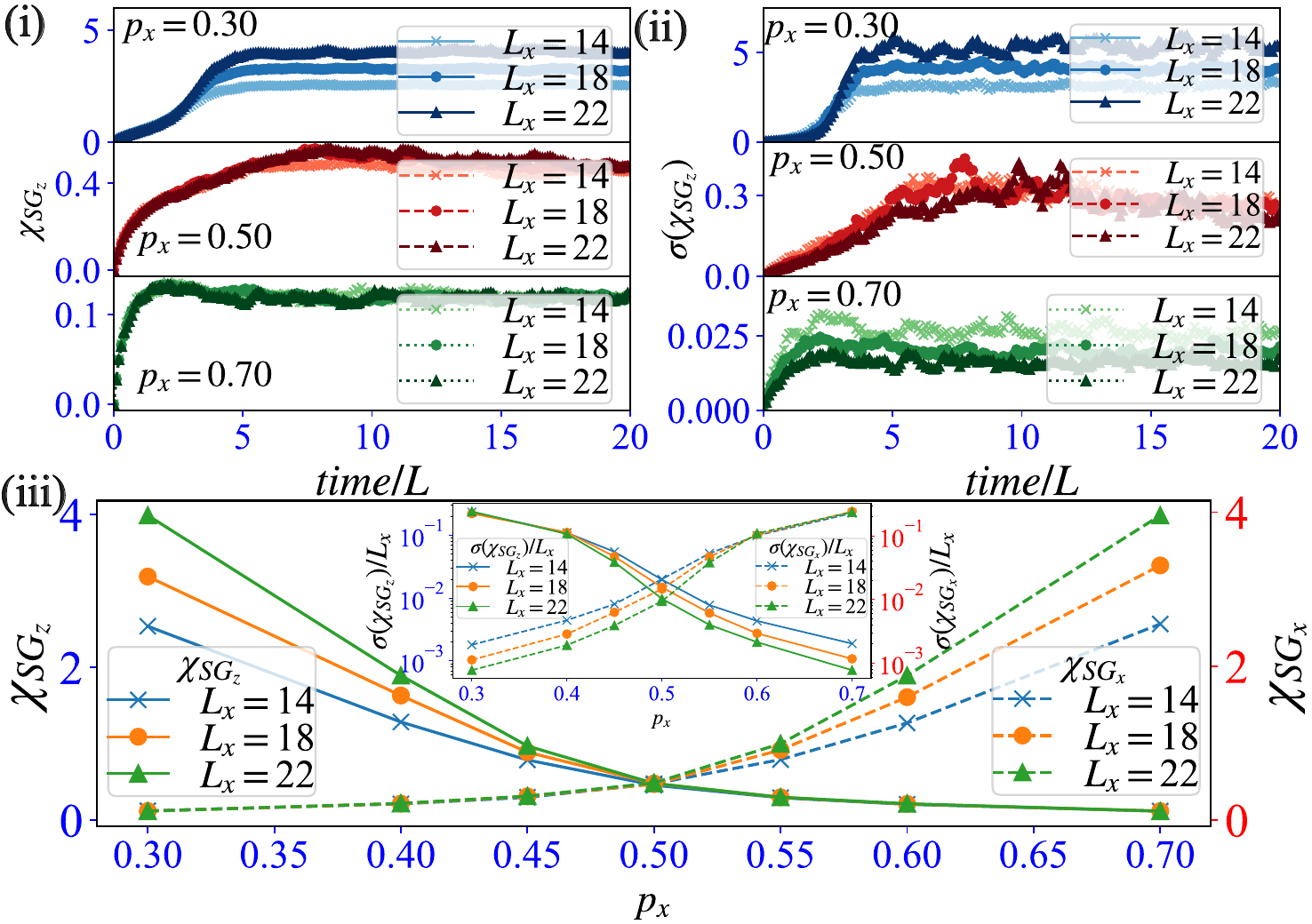}  
\end{center} 
\caption{(i) Long-time dynamics of Edward-Anderson spin-glass (SG) order parameter for $p_s=0.5$ and various values of $p_x=1-p_z$.
The SG order exhibits different behavior depending on $p_x$.
(ii) Sample-to-sample variance for the SG order. (iii) The late time saturation value of the SG order.
The inset of (iii) shows $\sigma(\chi_{SG_{x(z)}})/L_x$, and we used the same data as in (iii).
We took both sample and time average. 
As for the time average, we employ the last nine data points.
We used $2000$, $2000$, and $1000$ samples for $L_x=14,18$ and $22$, respectively.}
\label{Fig3}
\end{figure*}
The projective measurement of layer 2 can be regarded as a recovery process.
As explained in \cite{Botzung}, the projective measurement of $\tilde{A}_v$($\tilde{B}_p$) makes $\tilde{A}_v$($\tilde{B}_p$) an element of stabilizer generator again by removing local $\sigma^{x(z)}$-stabilizer generator if it has been produced by a layer 1 measurement.
Thus, we expect that for the case with  $p_s \gg p_{x}, p_z$, the recovery process succeeds in sustaining the TO of the initial state.
On the other hand for $p_s \ll p_{x(z)}$, stationary states acquire properties of Higgs or confined phase from the viewpoint of the gauge-Higgs model.
In particular, in the Hamiltonian picture for Higgs phase (the large-$J_z$ limit), $\sigma^z_{\ell}$'s in the bulk get expectation value such as $\sigma^z_{\ell}=-1$ (Higgs condensate), and then the model $H_{\rm TC}$ induces an effective transverse field Ising model \cite{Verresen2022} for the degrees of freedom on the upper rough boundary. 
Thus, it is expected that spontaneous symmetry breaking (SSB) of $Z_2$ symmetry takes place on the boundary leading to a magnetic LRO of $\sigma^z$. 
A counterpart (dual) picture to the above holds in the confined phase for large $p_x$ because of the electric-magnetic duality, an exact symmetry of the present MoC. 
A long-range order of $\sigma^x$ emerges on the smooth boundary as a signal of the SPT-confined phase.
Beyond the above intuitive pictures, the intermediate regime $p_x\sim p_z \sim 0.5$ does not have the magnetic-one-form symmetry nor electric one, and therefore, a sharp SSB of both $Z_2$ symmetries might not be observed there.
However, detailed Monte Carlo simulations of $H_{\rm TC}$ \cite{Gliozzi2006,Tupitsyn2010,Wu2012} have discovered a very interesting \textit{first-order} phase transition line $h_x=h_z$ in the very vicinity of TO transition, which corresponds to $p_x=p_z=0.5$ in the present MoC. 
This point will be discussed after showing numerical calculations of physical quantities.

In the rest of this study, we investigate the above qualitative picture in detail, especially critical regimes separating three phases. 
The measurement protocol is numerically performed by the efficient numerical simulation by the stabilizer formalism \cite{Gottesman1997,Aaronson2004}. 
Throughout this study, we ignore the sign and imaginary factors of stabilizers, which give no effects on observables that we focus on.

\begin{figure}[t]
\begin{center} 
\includegraphics[width=6cm]{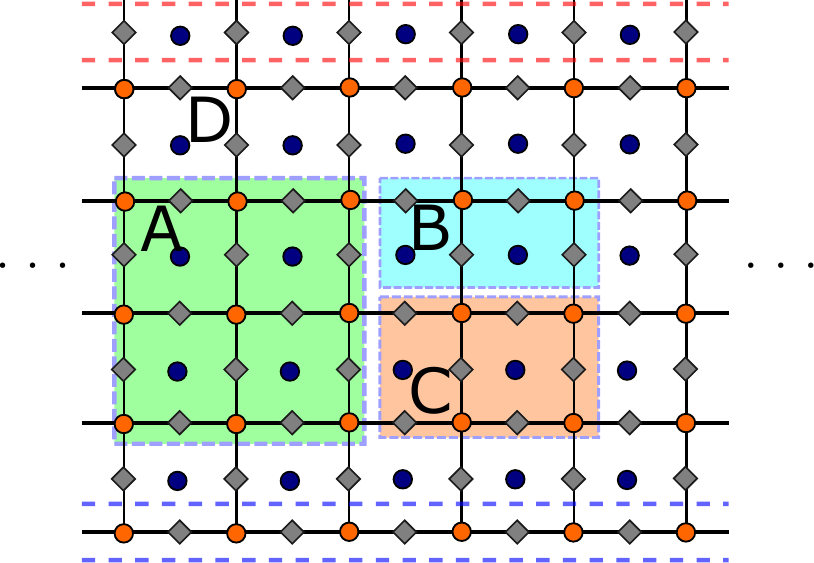}  
\end{center} 
\caption{Schematic image of the partition pattern of the system for calculating  topological entanglement entropy.}
\label{Fig4}
\end{figure}

This work clarifies the following problems in the MoC;
(i) how and when boundary LROs emerge on the upper rough and lower smooth boundaries 
(ii) how and when the bulk TO disappears.
In the numerical study addressing the above problems, we fix $L_y=10$ and vary $L_x$ up to 26.
For the first problem, one may expect that realizing LROs can be identified by a standard long-rand order parameter, e.g., $\langle\psi| \sigma^{z(x)}_{\ell_x}\sigma^{z(x)}_{\ell'_x}|\psi\rangle-\langle\psi| \sigma^{z(x)}_{\ell_x}|\psi\rangle\langle\psi|\sigma^{z(x)}_{\ell'_x}|\psi\rangle$; however, this is not the case.
This is because realized LROs due to measurement are more likely to be a glassy-type LRO, which closely relates to each measurement outcome. The glassy-type LRO is to be measured by an order parameter that is independent of  direction of 
outcome `spins' and merely depends on the magnitude of their correlations. 
Thus, to evaluate the glassy type LRO, we employ the Edward-Anderson SG order parameter \cite{Sang2021} defined by
\begin{eqnarray}
\chi_{SG}=\frac{1}{L_x}\sum_{\ell_x,\ell'_x}C_{SG}(\ell_x,\ell'_x),
\end{eqnarray}
with
$C_{SG}(\ell_x,\ell'_x)=\langle \psi|\sigma^z_{\ell_x}\sigma^z_{\ell'_x}|\psi\rangle^2-\langle \psi|\sigma^z_{\ell_x}|\psi\rangle^2 \langle \psi|\sigma^z_{\ell'_x}|\psi\rangle^2$,
where $|\psi\rangle$ is a stabilizer state and $\ell_x$'s are links on the boundary. 
As a target observable, we observe the variance of $\chi_{SG}$ divided by $L_x$, $\sigma(\chi_{SG})/L_x$.
This quantity measures sample-to-sample fluctuations and is useful to search a phase transition on the boundary.
For the second problem, we calculate topological entanglement entropy (TEE), $\gamma$.
In the present study, we employ the partition of the system shown in Fig.~\ref{Fig4} \cite{Kitaev_Preskill}, 
and for that partition, $\gamma$ is defined as 
\begin{eqnarray}
\gamma &=&S_A+S_B+S_C-S_{AB}-S_{BC}-S_{AC}+S_{ABC},\nonumber
\end{eqnarray}
where $S_{A}, S_{B}, \cdots$ denote entanglement entropy of the corresponding subsystem, which can be calculated from the number of linearly-independent stabilizers within a target subsystem and the number of qubit of the subsystem \cite{Fattal2004,Nahum2017}.  
The technical details of the methods to calculate EE are given in Appendix A in Ref.~\cite{KOI2023} and Appendix E in Ref.~\cite{Gullans2020}. 
Here, we employ EE with base-2 logarithm, following to the conventional definition of quantum information. As one example, the EE for $S_A$ is given by $S_A=-\operatorname{tr}[\rho_A \log_2 \rho_A]$ where $\rho_A$ is reduced density matrix for A-subsystem.
If the system is an exact topologically-ordered state, we have $\gamma=-1$. 
The TEE is a useful and standard quantity for investigating bulk TO in the MoC.
In particular, even for one of topological degenerate ground states, the TEE works efficiently to observe the TO.
In addition, we consider another partition for TEE \cite{Levin_Wen}, the numerical results of which are given in Appendix C.

\section{Dynamics of boundary order parameter}
We first observe long-time dynamics of Edward-Anderson SG order for $p_s=0.5$ and various values of $p_x$. 
For relatively small $p_x=0.3$ $(p_z=0.7)$, the dynamics of $\chi_{SG_z}$ is displayed in Fig.~\ref{Fig3}(i). 
The value of $\chi_{SG_z}$ increases to saturate $\mathcal{O}(1)$, signaling a long-range SG order on the upper rough boundary corresponding to the Higgs=SPT phase.
There, we also observe large fluctuations as in Fig.~\ref{Fig3}(ii). 
From the viewpoint of the Hamiltonian in Eq.~(\ref{HTC}), the leading term of the present protocol is the $J_z$-term as $p_z>p_s,p_x$, which is nothing but the charged Higgs particle hopping. 
Then, the coherent Higgs condensation, accompanying Wilson loop condensate, takes place that breaks the TO of the toric code, by the restoration of the magnetic one-form symmetry. 
The recent studies \cite{Verresen2022} showed that this symmetry restoration makes the Higgs phase one of SPTs, which is recognized by observing the degeneracy of the states on the rough boundary, the signal of which has been already verified by numerical methods \cite{KOI2024,Sukeno2024}. 
The above finite values of $\chi_{SG_z}$ come from the SPT, and then the bulk state is named Higgs=SPT. 
In passing, as one of the qualitative explanations, another type of symmetry is to be considered \cite{KOI2024}, the brief explanation of which is given in Appendix A. 

Next for $p_s=0.5$ and $p_x=0.7(p_z=0.3)$, $\chi_{SG_z}$ in Fig.~\ref{Fig3}(i) just exhibits a tiny increase and its fluctuations are very small [Fig.~\ref{Fig3}(ii)].
This result is consistent with the expectation that the Higgs condensate is suppressed for small $p_z$. 
Electric-magnetic duality indicates that $\chi_{SG_x}$ on the lower smooth boundary exhibits almost the same behavior with $\chi_{SG_z}$ on the upper rough boundary for $p_z=0.7$, and we verified this numerically [not shown].
Therefore, condensation of magnetic flux takes place in this parameter region producing the confined phase. 

Figure \ref{Fig3}(i) shows that the case with $p_s=0.5$ and $p_x=0.5$ is in between. 
This regime is simply featureless without any orders. 
Later study on the TO in the deconfined (toric code) phase will verify this expectation.
We also verified that the stable deconfined (toric code) phase emerges for relatively large $p_s$ as a result of the recovery effects.

Finally in Fig.~\ref{Fig3}(iii), we observe $p_x$-dependence of the late time value of $\chi_{SG_z}$ and $\chi_{SG_x}$ for $p_s=0.5$.
For small $p_x$, the late time SG order is enhanced. 
Close look at behavior of the SG order signifies that a crossover, instead of a phase transition, takes place, i.e., the late time values of $\chi_{SGz} (\chi_{SGx})$ smoothly increase with decreasing (increasing) $p_x$.
To ensure if the phase transition exists, we further focus on the variances of $\chi_{SGz}$ and $\chi_{SGx}$. The inset of Fig.~\ref{Fig3}(iii) shows the variances of $\chi_{SGz}$ and $\chi_{SGx}$, and we found that these two quantities do not show any significant behavior, i.e., monotonic increase or decrease.
This result is in contrast to the observation for the pure $\sigma^z$-measurement protocol corresponding to the deconfined (toric code)-Higgs transition studied in the previous paper Ref.~\cite{KOI2024}, where $\sigma(\chi_{SG_z})/L_x$ shows a sharp peak, and its peak becomes more significant as $L_x$ increases, implying the existence of the conventional second-order phase transition. 
In the present case, however, no quantities are indicating a bulk thermodynamic phase transition for the Higgs-confined phase.
We expect that this reflects to the behavior of the boundary SG order indicating a novel bulk-boundary correspondence. 
We explained in the previous study~\cite{KOI2024} that a finite expectation value of open Wilson loops (EVOWL) generates effective spin couplings on the rough boundary and the SG order emerges as a result. 
On the deconfined (toric code)-Higgs phase transition, the EVWL (more precisely, Fredenhagen-Marcu string order) exhibits a transition-like behavior similar to the magnetization in the (2+1)D Ising model, whereas in the confined-Higgs crossover regime, both the Wilson and 't Hooft string symmetries do not emerge as Fradkin-Shenker observation dictates (see also \cite{Xu2024}).
Reflecting this smooth change in the bulk properties, the SG order exhibits crossover-like behavior
instead of a genuine transition as we observed numerically.
This observation obtained by the numerical study on the boundary SG order is complementary 
to the work in Ref.~\cite{Verresen2022} (in which the boundary transition between Higgs-confined regimes was investigated by density-matrix-renormalization-group methods).
\section{Dynamics of topological entanglement entropy}
\begin{figure}[t]
\begin{center} 
\includegraphics[width=8.5cm]{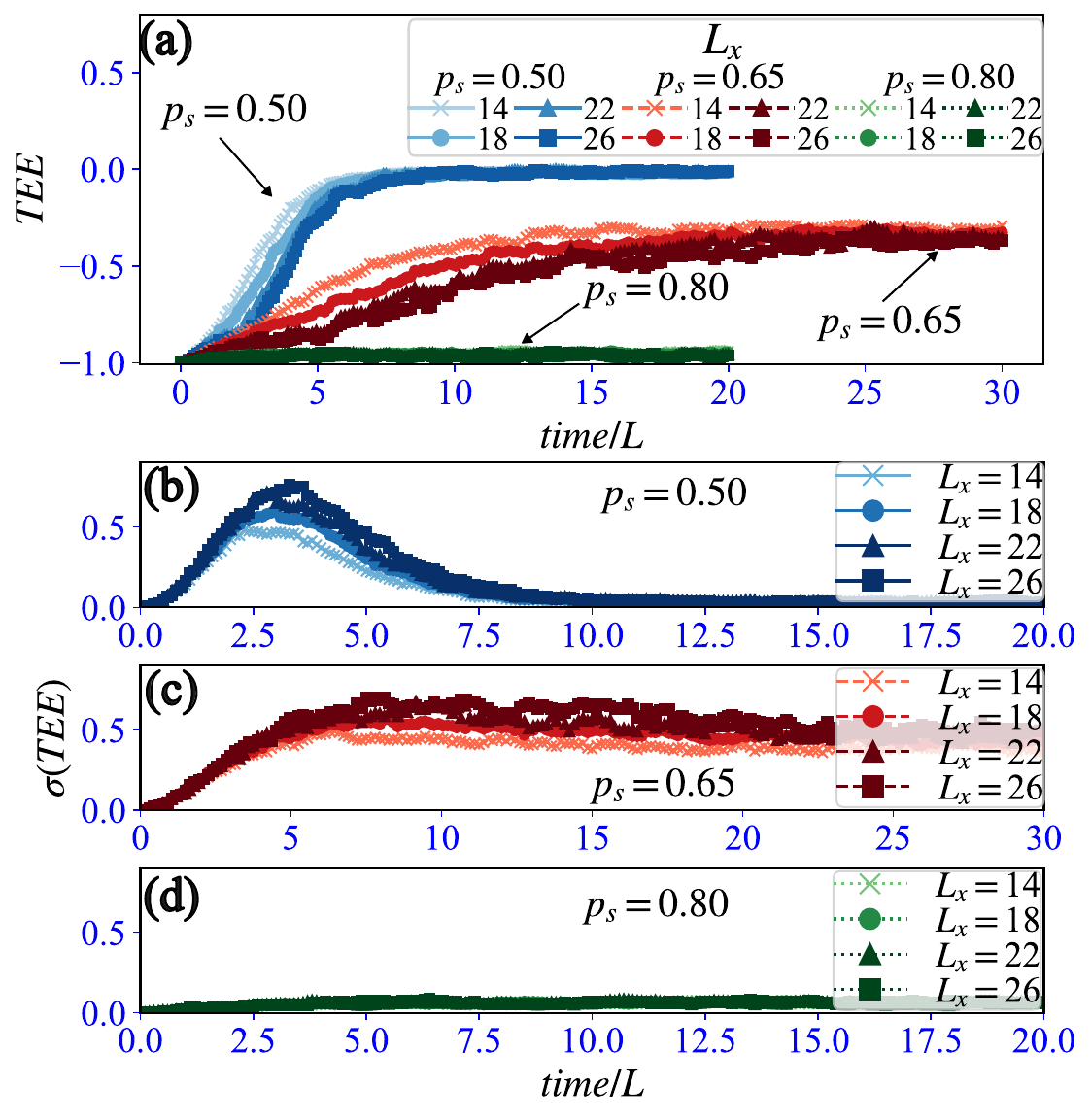}  
\end{center} 
\caption{(a) Time evolution of topological entanglement entropy $\gamma$ for $p_x=p_z=0.5$ and 
various $p_s$ and system sizes.
(b) $\sim$ (d): Fluctuation of topological entanglement entropy.
We used $2000$, $2000$, $1000$ and $1000$ samples for $L_x=14,18,22$ and $26$, respectively.
}
\label{Fig5}
\end{figure}

We move on to the dynamics of the TEE. 
We first study the recovery effect by the measurement of layer 2 for a fixed probability of the measurement of layer 1. 
The numerical results for various sizes $L_x$ are displayed in Fig.~\ref{Fig5} for the most competitive case of layer 1, $p_x=p_z=0.5$.
For a large recovery case $p_s=0.80$, the TEE starts with $\gamma=-1$ and exhibits almost no changes in the long time evolution. 
The stationary state sustains TO in the bulk staying in the deconfined phase.
As shown in Fig.~\ref{Fig5} (d), in the whole time evolution, fluctuations are rather small. 
On the other hand for a relatively weak recovery case with  $p_s=0.50$, behavior of the TEE drastically changes from that of large recovery case as seen in Figs.~\ref{Fig5} (a) and (b).
The late time value of $\gamma$ saturates to $\gamma=0$, indicating that the stationary state loses the initial TO in the bulk. 
We also find that the fluctuation of the TEE is large in the regime where the TEE changes drastically.
This behavior of the TEE in this MoC obviously indicates dynamic phase transition from the deconfnied to Higgs phases.
In order to obtain a critical probability, $p^c_s$, for the transition, we further observe the late time value of the TEE around $t/L= 20$ (for $p_s=0.5,0.6,0.75,0.8,0.9$) and $t/L=30$ (for $p_s=0.65, 0.7$) and find $p^c_s\simeq 0.65$.
The results are displayed in Figs.~\ref{Fig5} (a) and (c).
There, the saturation value of $\gamma$ is fluctuating around $\gamma=-0.5$ and its fluctuation exhibits large values for a long period, exhibiting a small system-size dependence. 

\begin{figure}[t]
\begin{center} 
\includegraphics[width=8.5cm]{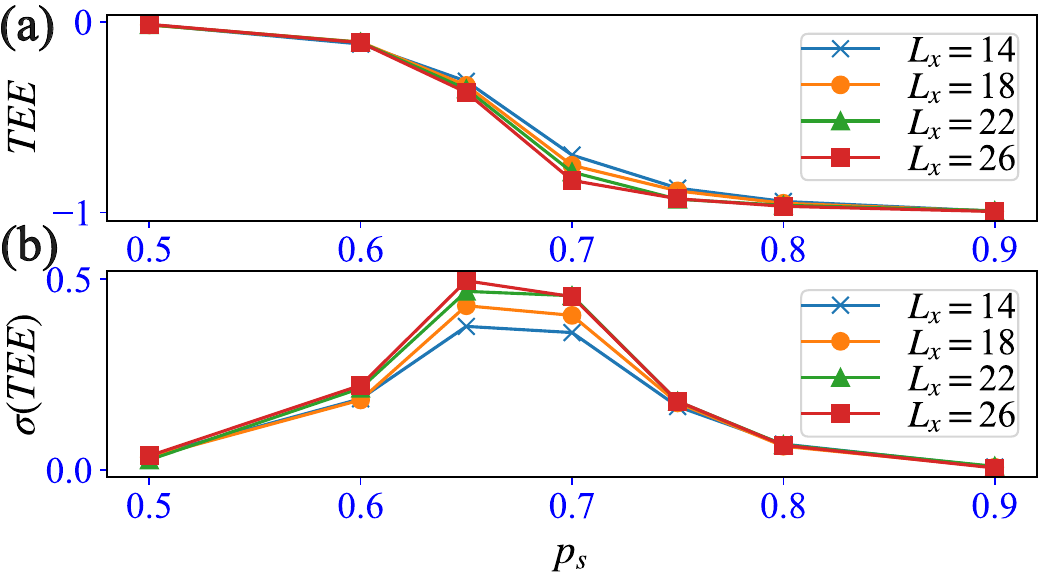}  
\end{center} 
\caption{Long time saturation values of topological entanglement entropy for $p_x=p_z=0.5$.
Topological entanglement entropy and its fluctuation indicate that the bulk phase transition takes place
at $p_s \simeq 0.65$.
Numerical calculations are performed up to $t=20L$ and $t=30L$ for $p_s=0.5,0.6,0.75,0.8,0.9$ and $p_s=0.65,0.70$, respectively.
We took both sample and time average. 
As for the time average, we employ the last nine data points.
We used $2000$, $2000$, $1000$, and $1000$ samples for $L_x=14,18,22$ and $26$, respectively.
}
\label{Fig6}
\end{figure}

Finally, we summarize the saturation vale of $\gamma$ as a function of $p_s$ for various system sizes in Fig.~\ref{Fig6}.
We find a clear step-function-like behavior of the TEE (panel (a)) and a peak of the fluctuations of the TEE is located at $p_s\simeq 0.65$ (panel (b)). 
We expect that in the thermodynamic limit, the saturation value of the TEE becomes a genuine step function of $p_s$. 
These results imply that there exists a clear bulk phase transition emerging by varying the recovery probability $p_s$, at which the bulk topological order vanishes.

We also investigate the behavior of the TEE for a non-contractible partition pattern of 
the system as in Ref.~\cite{Levin_Wen}, and obtain similar results to the above, which are shown in Appendix C.
The initial TC state has $\gamma=-2$ for the non-contractible partition, and the state changes its value depending on the probability rates. The critical probability is estimated as $0.65$ for $p_x=p_z=0.5$.

\section{Conclusion and discussion}
The present work clarified the dynamics of MoC, in which we performed competitive measurements on the TC state on a cylinder geometry with distinct upper and lower boundary conditions 
to observe the duality property between the Higgs and confined phases.
There, the measurement operators correspond to the terms in the Hamiltonian of the TC model in magnetic fields.
In this MoC, stationary states, which seem to have properties of Higgs, confined and critical phases, emerge by the `time evolution' from the exact stabilizer state of the deconfined (toric code) phase on the cylinder geometry. 
We identified physical properties of emergent states by numerically observing the boundary long-range magnetic orders and TEE in the bulk. 
We found that the critical measurement probability ratios corresponds to the ratios of the parameters in the lattice gauge-Higgs model Hamiltonian. 
We further investigated how the initial TC state develops under various settings of the measurement. 
In particular, we observed that the Higgs and confined phases are separated from the deconfined phase by topological entanglement entropy, whereas the Higgs and confined phases are distinguishable with
each other via a crossover of boundary modes.

\begin{figure}[t]
\begin{center} 
\includegraphics[width=8.5cm]{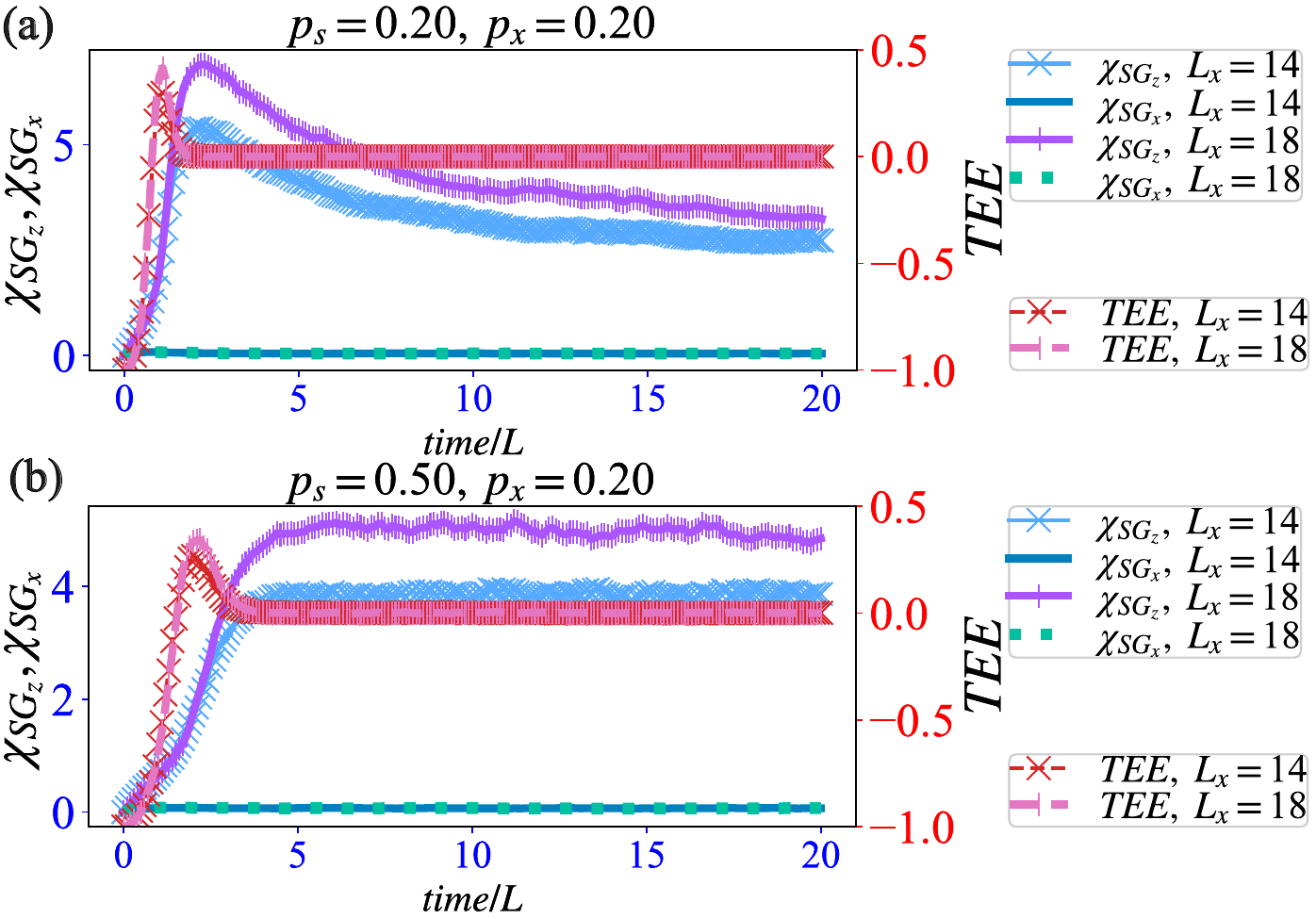}  
\end{center} 
\caption{Time evolution of the spin-glass (SG) order and topological entanglement entropy for $p_x=0.2$.
Data of both the cases show that the topological entanglement entropy tends to vanish before the development of the SG order.
This behavior can be interpreted that condensation of Wilson loop in the bulk is the origin of the SG order on the boundary.
The sample averages are taken over $1000$ samples for both $L_x=14$ and $L_x=18$.
}
\label{Fig7}
\end{figure}
In the present MoC, we observed simultaneously both the SG LROs on the upper-rough and lower-smooth boundaries and TEE in the bulk.
This gives us an important insight into the relation between the SPT and intrinsic TO in the gauge-Higgs models.
As shown in Fig.~\ref{Fig7}, the `time evolution' of the SG order on the boundaries and TEE in the bulk is observed clearly to find that the bulk topological transition takes place first and the SG transition follows that.
This behavior is observed generally.
As we explained in the main text, the SG order stems from the bulk one-form-symmetry, in particular, condensation of the Wilson ('t Hooft) string. 
This condensate induces effective Ising-type long-range couplings between spins on the boundaries producing
the SG order. 
Numerical study in Fig.~\ref{Fig7} can be interpreted naturally in such a way that a sufficiently large coupling between boundary spins is necessary for the SG order to emerge.
In fact, the recent study on Wilson string condensate for lattice gauge-Higgs model in Ref. \cite{Xu2024} shows that the condensate behaves as an order parameter and exhibits $(2+1)-D$ Ising spin criticality developing continuously from zero.
The difference in the location of the two transitions, which was observed already in our previous work~\cite{KOI2024}, supports the above observation concerning to the bulk-boundary correspondence. 


\section*{Acknowledgements}
This work is supported by JSPS KAKENHI: 
JP23KJ0360(T.O.) and JP23K13026(Y.K.). 

\appendix
\section*{Appendix A: Parameter ratio and probability ratio correspondence}
In the previous study by two of the authors of this work \cite{KI2023}, a conjecture was given; parameter ratio and probability ratio correspondence. 
The target Hamiltonian is the following,
\begin{eqnarray}
H_{\rm stab}=-\sum^{L-1}_{j=0}\sum^{M}_{\alpha=1}J^{\alpha}K^{\alpha}_{j},\nonumber
\label{Hstab2}
\end{eqnarray}
where $J^{\alpha}(>0)$ and $L$ is the total number of sites $\{ j\}$, $\alpha$ represents $M$-types of stabilizers anti-commuting with each other, i.e., 
$\{ K^{\alpha}_j\}$ satisfy $[K^{\alpha}_j,K^{\alpha}_k]=0$ and $(K^{\alpha}_j)^2=1$, and for different types of stabilizers, $[K^{\alpha}_j,K^{\beta}_k]\neq 0$ and $\{K^{\alpha}_j,K^{\beta}_k\}=0$ ($\alpha\neq \beta$).

To construct a corresponding MoC, we pick up a single stabilizer among the different types of $K^\alpha_{j_0}$ with a probability $p^{\alpha}$ and choose a target site $j_0$ with equal probability $1/L$ at each time step. 
We set the probability condition of the choice of the type of the stabilizer, such as $\sum_{\alpha}p^\alpha=1$. 
The setup is the same as that of the previous works \cite{Lavasani2021,Klocke2022}.

Then, the ground-state phase diagram of $H_{\rm stab}$ determined by the parameters $J^\alpha$ is very 
close to the phase diagram of the steady state in the MoC, which is determined by an ensemble average of the measurement pattern of the MoC. 
This fact indicates that the ratio of parameters $\frac{J^\alpha}{J^{\beta}}$ is related to the ratio of probabilities $\frac{p^{\alpha}}{p^{\beta}}$, that is, $\frac{J^\alpha}{J^{\beta}}\longleftrightarrow\frac{p^{\alpha}}{p^{\beta}}$. 
This relationship is nothing but the explicit form of ``parameter ratio-probability ratio correspondence''. 
The qualitative verification of the above consideration was analytically carried out in Ref.~\cite{KI2023}.\\

\section*{Appendix B: Non-local gauge invariant operator}
To qualitatively understand the boundary LRO in the Higgs phase in the toric code system with the upper rough and lower smooth boundaries on the cylinder geometry, 
we introduce non-local gauge invariant operator (NLGIO) symmetries for $H_{\rm TC}$ with $J_x=0$. 
Here, we focus on the upper rough boundary and Higgs phase with noticing that parallel argument is possible for the lower smooth boundary and confined phase.
Then, the symmetry has two types: 
(I) $G_{lo,1}\equiv \prod_{\ell\in\Gamma_0} \sigma^z_{\ell}$
with an arbitrary close loop $\Gamma_0$, 
(II) $G_{lo,2}=\sigma^z_{\ell_{r1}}\biggl[\prod_{\ell\in\Gamma_b}\sigma^z_{\ell}\biggr]\sigma^z_{\ell_{r2}}$ where $\ell_{r1}$ and $\ell_{r2}$ are two arbitrary dangling links on the upper rough boundary and a string in the bulk $\Gamma_b$ connecting the $\ell_{r1}$ and $\ell_{r2}$ links. 
\begin{figure}[t]
\begin{center} 
\vspace{1cm}
\includegraphics[width=6cm]{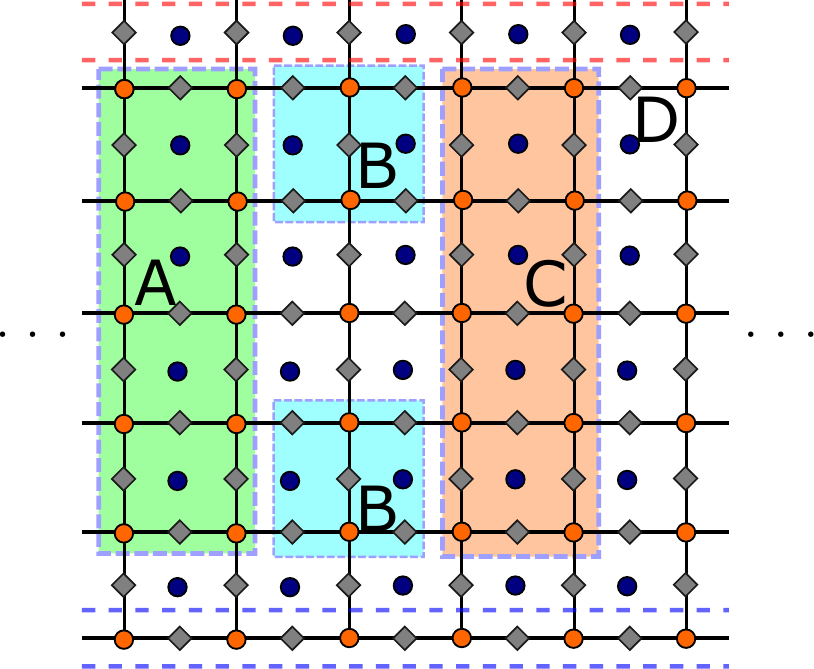}  
\end{center} 
\caption{Schematic image of a non-contractible partition pattern for topological entanglement entropy~\cite{Levin_Wen}.}
\label{Fig8}
\end{figure}
Both these operators satisfy $[H_{\rm TC},G_{lo,1(2)}]=0$ for $J_x=0$.
The second-type NLGIO $G_{lo,2}$ plays a key role to understand the boundary LRO through the Higgs condensation. 
That is, the Higgs phase can be regarded as a symmetry-restored state of the second-type NLGIO $G_{lo,2}$.
\begin{figure}[t]
\begin{center} 
\vspace{1cm}
\includegraphics[width=8cm]{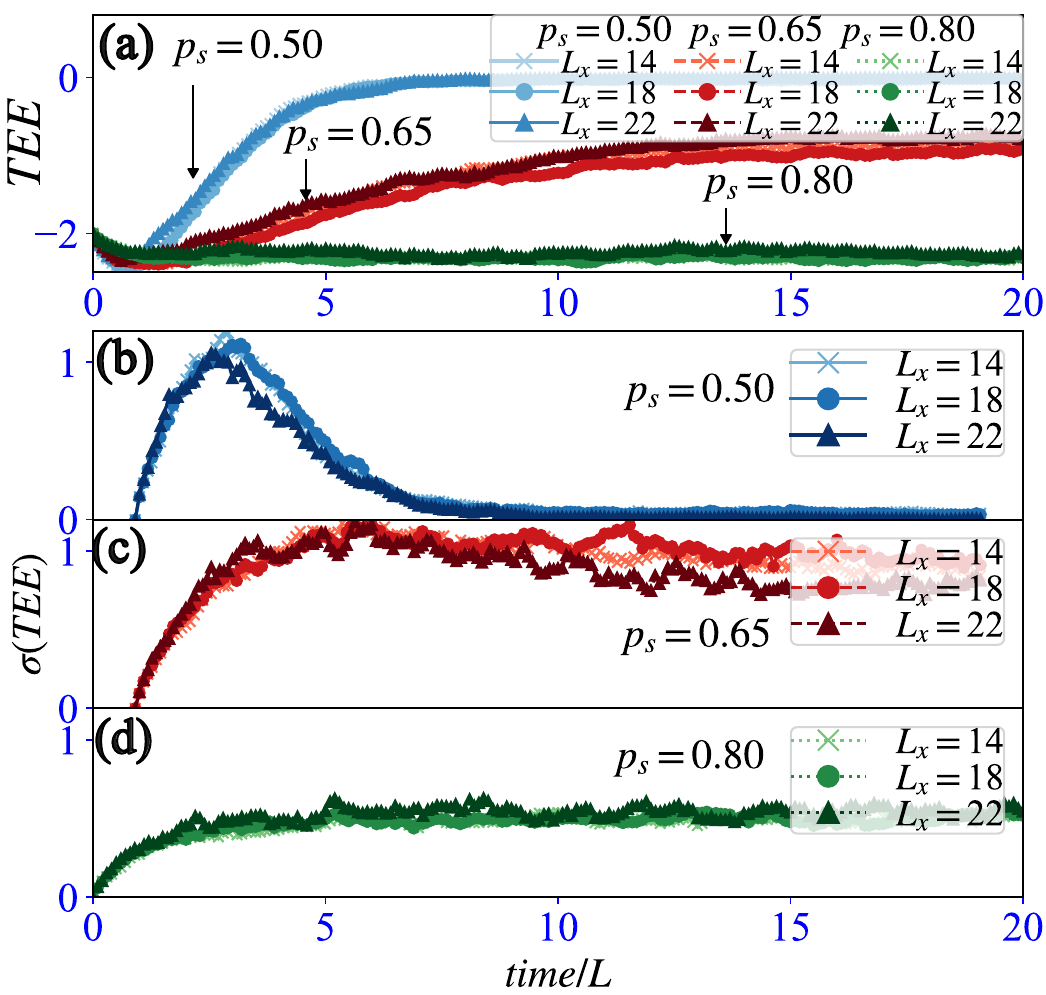}  
\end{center} 
\vspace{-0.5cm}
\caption{Time evolution of topological entanglement entropy~\cite{Levin_Wen} for $p_x=p_z=0.5$ and various $ps$ and system sizes $L_x$. We fixed $L_y=10$.
Panels (b)-(d) represent fluctuations of TEE for (b) $p_s=0.50$, (b) $p_s=0.65$, and (c) $p_s=0.80$, respectively.
The sample averages are taken over $1000$, $1000$, and $500$ samples for both $L_x=14,18$ and $22$, respectively.
}
\label{Fig9}
\end{figure}
In the deep Higgs phase, the local $\sigma^z_\ell$ stabilizer generator is proliferated in the bulk, leading to a finite string order, $\langle \prod_{\Gamma_b} \sigma^z_\ell\rangle\neq 0$ in the system, regarded as the Higgs condensation. 
As the state of the Higgs phase  $|\psi_{\rm Higgs}\rangle$ respects the NLGIO symmetry, then the following relation is satisfied, $G_{lo,2}|\psi_{\rm Higgs}\rangle=\sigma^z_{\ell_{r1}}\biggl[\prod_{\ell\in\Gamma_b}\sigma^z_{\ell}\biggr]\sigma^z_{\ell_{r2}}|\psi_{\rm Higgs}\rangle \propto |\psi_{\rm Higgs}\rangle$.
That is, the Higgs condensation in the bulk gives the result $\langle \sigma^z_{\ell_{r1}}\sigma^z_{\ell_{r2}}\rangle\neq 0$,
the emergence of the LRO on the rough boundary.
We expect that the NLGIO symmetries have a similar robustness to the general one-form-symmetry,
and the above observation holds for a finite-$J_x$ system as long as it belongs to the Higgs phase.
Numerical studies in the present work support this expectation.

We note again that the above discussion of the NLGIO symmetry can be applicable to the emergence of the LRO of $\sigma^x_\ell$ 
on the smooth boundary in the confinement phase by the dual picture.

\section*{Appendix C: topological entanglement entropy of another partition}

In this Appendix, we show numerical calculations of the TEE employing another type of system partition pattern in Fig.~\ref{Fig8}. 
The TEE, $\gamma$, is given by 
\begin{eqnarray}
\gamma = S_A+S_B+S_C-S_{AB}-S_{BC}-S_{AC}+S_{ABC},\nonumber
\end{eqnarray}
and $\gamma=-2$ for the genuine TC state under this partition and vanishes for a topologically-trivial state.
In this pattern of the partition, we are afraid that the TEE exhibits less clear time-evolution behavior than that of in Fig.~\ref{Fig5} as the region D is divided into two distinct portions.
However, this investigation supports the observation of the TEE obtained for the pattern in Fig.~\ref{Fig5},
in particular, the location of the phase transition.

Numerical results for $p_x=p_z=0.5$ are displayed in Fig.~\ref{Fig9} which indicate the critical probability $p^c_s\simeq 0.65$, in good agreement with the result in Fig.~\ref{Fig5}.


\end{document}